\documentclass[5p,12pt]{elsarticle}

\usepackage{amssymb}
\usepackage{graphics}
\usepackage{graphicx}
\usepackage{amssymb}
\usepackage{epstopdf}
\usepackage{dcolumn}
\usepackage{color}
\def\BaOsS {Ba$_2$NaOsO$_6$ }
\def\BaOs  {Ba$_2$NaOsO$_6$}
  \def \Na {$^{23}$Na }

   \def \ie {{\it i.e.} }

\journal{Physica B: Condensed Matter - SI:PHYSB SCES 2017}

\begin{document}

\begin{frontmatter}



\title{Phase Diagram of Ba$_{2}$NaOsO$_{6}$, a Mott insulator with strong spin orbit interactions}


\author{W. Liu}
\author{R. Cong}
\author{E. Garcia}
\address{Department of Physics, Brown University, Providence, RI 02912, U.S.A.}
\author{A. P. Reyes}
\address{National High Magnetic Field Laboratory, Tallahassee, FL 32310, U.S.A.}
\author{H. O. Lee}
\address{Department of Applied Physics and Geballe Laboratory for Advanced Materials, Stanford University, California 94305, U.S.A}
\author{I. R. Fisher}
\address{Department of Applied Physics and Geballe Laboratory for Advanced Materials, Stanford University, California 94305, U.S.A}

\author{V. F. Mitrovi{\'c}\corref{cor1}}\ead{vemi@brown.edu}

 \address{Department of Physics, Brown University, Providence, RI 02912, U.S.A.}

 \cortext[cor1]{Corresponding author}

\begin{abstract}
We report  \Na nuclear magnetic resonance (NMR)  measurements of  the Mott insulator with strong spin-orbit interaction \BaOsS    as a  function of    temperature in different magnetic fields ranging from 7 T to 29 T.  The measurements, intended to  concurrently probe spin and orbital/lattice degrees of freedom,  are an extension of our work at lower fields reported in \cite{Lu17}.  We have identified clear quantitative NMR signatures that display the appearance of a canted 
 ferromagnetic phase,  which is  preceded by local point symmetry breaking. We have compiled the field temperature phase diagram extending up to \mbox{29 T}. We find that the broken local point symmetry phase extends over a wider temperature range as  
 magnetic field increases.
\end{abstract}


\end{frontmatter}



\section{Introduction}
\label{Intro}
The combined effects of strong electronic correlations with spin-orbit coupling (SOC) can lead to a plethora of emergent novel quantum states. In the quest to predict and identify the emergent phases, studies have focused on  magnetic Mott insulators with strong SOC  \cite{Jackeli09, ChenBalents10, ChenBalents11, KrempaRev14}. Strong SOC can significantly enhance quantum fluctuations and amplify effects of frustration leading to novel quantum states \cite{ChenBalents10,ChenBalents11,Dodds11}. 
 A key feature of  theoretical models for predicting the emergent properties  is 
  that significant interactions are  fourth and sixth order in the effective spins, due to    strongly orbital-dependent exchange \cite{ChenBalents10}. 

The Mott insulating $d^{1}$ double perovskites with cubic structure are model systems of  Mott insulators with strong SOC.  An example material is \BaOs,    a double perovskite with Na and Os ions inhabiting alternate cation ÒBÓ sites, which for an undistorted structure has a  face-centered-cubic lattice. 
We have performed microscopic measurements  designed to concurrently probe spin and orbital/lattice degrees of freedom in this compound and so provide stringent tests of theoretical approaches.  
Our static NMR measurements   reveal that the local cubic symmetry breaking, induced by a deformation of the oxygen octahedra, precedes the formation of the long range ordered magnetism \cite{Lu17}.  Specifically, we found that these deformations generate an orthorhombic point symmetry in the magnetic phase. Furthermore, 
  we established that the magnetically ordered state is the  canted two-sublattice ferromagnet (FM), believed to be driven by the staggered quadrupolar order \cite{ChenBalents10}.  Our observation of both the local cubic symmetry breaking and appearance of  two-sublattice  exotic FM phase is in line with theoretical predictions based on quantum models with  multipolar magnetic interactions  \cite{ChenBalents10}.  Thus,   our findings establish that such quantum models  with  multipolar magnetic interactions  represent an appropriate theoretical framework for predicting emergent properties in  materials with both strong correlations and SOC, in general. To provide  further tests of quantum models, here we present the extension of   our work in \mbox{Ref. \cite{Lu17}} to high magnetic fields.   
We  compile the field temperature phase diagram extending up to \mbox{29 T}. We find that as 
 magnetic field increases, broken local point symmetry  phase extends over wider temperature range. 

\section{Experimental Technique and the Sample}
\label{Exp}

The measurements were done   at the NHMFL in Tallahassee, FL    
using both high homogeneity superconducting and resistive magnets.   The temperature control was provided by $^4$He variable temperature  insert. 
The NMR data was recorded using a state-of-the-art laboratory-made NMR spectrometer.
The spectra were obtained, at each given value of the applied field, from the sum of spin-echo Fourier transforms recorded at constant frequency intervals. 
We used a standard spin echo sequence $(\pi/2-\tau-\pi)$.    A gyromagnetic ratio of  $^{23}\gamma$ = 11.2625 MHz/T  was used for all frequency to field scale conversions.
For nuclear sites with spin  $I  >   1/2$, such as $^{23}$Na with $I=3/2$, and non-zero   electric field gradient (EFG),   quadrupole interaction between nuclear spin and EFG splits otherwise single NMR line to $2I$ lines \cite{AbragamBook}. Thus,  the $^{23}$Na spectral line splits in three in the presence of a non-zero EFG.  However, for small finite values of the EFG  three peaks are not necessarily discernible,  in which case  significant line broadening can only be detected. 
 At  sites with  cubic point   symmetry the EFG is zero, as is the case for Na nuclei in the high temperature PM phase, and a single line is observed.

  The sample was both zero-field and field-cooled. We did not   detect any influence of the sample's cooling history on the NMR spectra.  
 Nevertheless for consistency, all results presented in the paper were obtain in field-cooled conditions. 
The sample was mounted to
one of the crystal faces and rotated with respect to the applied field about an axis using a single axis goniometer. The direction perpendicular to the face to which the sample was mounted denotes [001] direction.  The labeling is only relevant in the low temperature magnetically ordered phase characterized by orthorhombic point symmetry.

 High quality single crystals of \BaOsS  with a truncated octahedral morphology were grown from a molten hydroxide flux, as described in  Refs. \cite{Stitzer2002, Erickson07}. Crystal quality was checked by x-ray diffraction, using a Bruker Smart Apex CCD diffractometer, which indicated that the room temperature structure belongs to the  $Fm\bar 3m$ space group  \cite{Erickson07}. NMR measurements were performed for a single crystal with a volume of approximately 1 mm$^3$. Such a small sample volume precludes effective neutron scattering studies of magnetic order. 
 The quality of the sample was confirmed by the sharpness of
    $^{23}$Na  NMR spectra both in the high temperature paramagnetic state and low temperature quadrupolar split spectra.

\section{Temperature and Field Evolution of NMR Spectra}

\begin{center}
 %
 %
\begin{figure}[t]
  \vspace*{-0.0cm}
\begin{minipage}{0.98\hsize}
 \centerline{\includegraphics[scale=0.69]{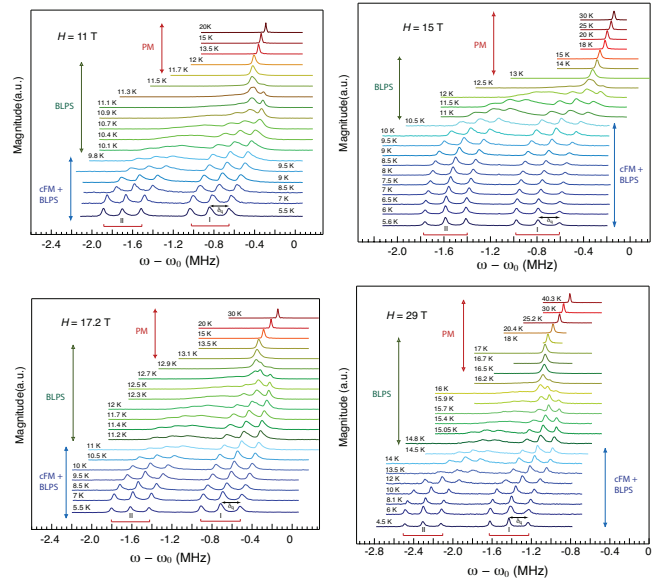}} 
\begin{minipage}{.98\hsize}
 \vspace*{-0.0cm}
\caption[]{\label{Fig1} \small 
Temperature evolution of $^{23}$Na spectra  
at  various strengths of magnetic field,   applied parallel to [001] crystalline axis, as denoted.   
 Narrow single peak spectra characterize high temperature paramagnetic (PM) state.  At intermediate temperatures, broader and more complex spectra reveal the appearance of an electric field gradient (EFG) induced by breaking  of local cubic symmetry, as described in \cite{Lu17}. 
Splitting into 2 sets of triplet lines (labeled as I and II),   reflecting the existence of two distinct magnetic sites in the lattice, is evident at lower temperatures. 
Zero of frequency is defined as $\omega_{0} =  \, ^{23}\gamma \,H$. Splitting between quadrupolar  satellites is  denoted by $\delta_{q}$.  Abbreviation: PM, paramagnetic; BLPS, broken local point (cubic) symmetry; and, cFM,  canted ferromagnetic.}
 \vspace*{-0.3cm}
\end{minipage}
\end{minipage}
\end{figure}
%
\end{center}

To determine  the field-temperature phase diagram of \BaOs,  we inspect the temperature $(T)$ dependence of 
 $^{23}$Na NMR spectra measured in applied magnetic fields ($H$) that exceed  those presented in \mbox{Ref. \cite{Lu17}}. 
 In  \mbox{Fig. \ref{Fig1}}, we plot the temperature evolution of   $^{23}$Na NMR spectra measured at four different magnetic fields ranging from 11 T to 29 T. The analysis of these  $^{23}$Na spectra allows us to determine the transition temperature $(T_{c})$ from paramagnetic   to low temperature ferromagnetic   state and the onset temperature for breaking of local cubic symmetry $(T^{*})$, as described in sections below.
 This is because these spectra reflect   the distribution of the hyperfine fields and the electronic charge and are thus a sensitive probe of both the  electronic spin polarization (local magnetism)  and charge distribution (orbital order and/or lattice symmetry).

%
  \vspace*{-0.00cm}
%
 %

%

High temperature   spectra consist of a single narrow NMR line. Since  the nuclear spin $I$ of $^{23}$Na equals to $3/2$, the absence of the three distinct quadrupolar satellite lines indicates that EFG is zero as a consequence of  a  cubic environment. Therefore, the observed  single narrow NMR spectra is evidence that the high temperature phase of \BaOsS   is  a  PM state characterized by cubic symmetry.   
On lowering the temperature, the NMR line broadens and splits into multiple peaks indicating onset of significant changes in  
 the local symmetry, thereby producing EFG, \ie asymmetric (non-cubic) charge distribution. 
 Therefore, the observed line broadening and subsequent splitting of the Na spectra into triplets, in the magnetically    ordered phase, indicates breaking of the  cubic point symmetry  caused by local distortions of electronic charge distribution, as established in \cite{Lu17}. 
 These distortions,   marking the broken local point symmetry (BLPS) phase, occur above  the transition into the magnetic state.  
For example, at low temperatures below 10 K at 11 T, the  \Na spectra clearly split into 6 peaks, that is two sets of triplet lines,   labeled  as I and II in \mbox{Fig. \ref{Fig1}}, that are well separated in frequency.  
 The emergence  of these two sets of  triplets  indicates the appearance of  two distinct magnetic sites, \ie two nuclear sites that 
  sense two different local fields   in the lattice, as described in \cite{Lu17}.  More precisely, detailed analysis of the NMR spectra as a function of the strength and direction of the applied field revealed that the  two sets of  triplets  originate from two-sublattice  canted ferromagnetic  phase \cite{Lu17}. Therefore, the two sets of triplet lines indicate the presence of the long range ordered (LRO) magnetism.
  
  Having described the main features of the NMR spectra, we proceed to deduce $H-T$ phase diagram, \ie transition temperatures from PM to low temperature  FM and BLPS phases.  
      
    \begin{center}	
 %
\begin{figure}[t]
  \vspace*{-0.0cm}
\begin{minipage}{0.98\hsize}
 \centerline{\includegraphics[scale=0.50]{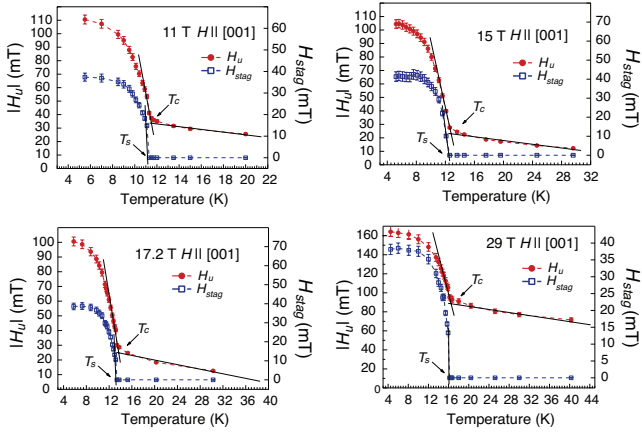}} 
\begin{minipage}{.98\hsize}
 \vspace*{-0.0cm}
\caption[]{\label{Fig2} \small 
Temperature dependence of the uniform and staggered fields measured 
at  various  magnetic fields    applied parallel to [001] crystalline axis.  
The uniform field data is denoted by filled circles while open squares represent staggered fields. 
Arrows mark transition temperature $(T_{c})$ from PM to low temperature magnetic state, determined by the crossing points of the plotted solid lines. These $T_{c}$ values are displayed in \mbox{Tab. \ref{Table1}}. 
Solid lines are linear fits to the data, while dashed lines are guide to the eyes. 
 $T_{s}$ marks the onset temperature for the appearance of the local staggered field.  
  }
 \vspace*{-0.1cm}
\end{minipage}
\end{minipage}
\end{figure}
%
\end{center}
  \vspace*{-0.00cm}
%

  %
%

\subsection{Low  Temperature - Magnetic Phase}

To determine the transition temperature $(T_{c})$ from paramagnetic  to low temperature  FM state,
we examine the temperature dependence of the local uniform $(H_{u})$ and staggered $(H_{stag})$ fields. 
We define  the local uniform $(H_{ u} = {1 \over 2} \left[ \langle H_{\rm I} \rangle +   \langle H_{\rm II} \rangle \right ])$ and staggered $(H_{stag}= {1 \over 2} \left[ \langle H_{\rm I} \rangle -   \langle H_{\rm II} \rangle \right ] )$ fields, where   
  the average is taken over the triplet I and II, as denoted.       
    We note that $H_{u}$ corresponds to the local field as determined by the first moment of the entire spectra. 

 \begin{table}[h]
\begin{center}
\begin{tabular}{|c|c|c|}\hline $\quad H \, \|  [001] \quad$ & \, $T_{c}$  & \, $T_{s}$   \\\hline 
 $0 \, {\rm T}$ & \, $\,6.3 \,  {\rm K}\,$ \cite{{Erickson07}}& \,      
  \\\hline $7 \, {\rm T}$ & \, $\,9.9 \,  {\rm K}\,$& \, $10.2 \,  {\rm K}  \,$     \\\hline 
$9 \, {\rm T}$ &  \,  $\,10.8\, {\rm K}\,$& \, $10.6 \,  {\rm K}  \,$    \\\hline 
$11 \, {\rm T}$ &  \, $\,11.6  \, {\rm K}\,$& \, $11.3 \,  {\rm K}  \,$  \\\hline 
$15 \, {\rm T}$ &  \, $\,13   \, {\rm K}\,$& \, $12.7 \,  {\rm K}  \,$  \\\hline 
$17.2 \, {\rm T}$ &  \, $\,13.6  \, {\rm K}\,$& \, $13.4 \,  {\rm K}  \,$  \\\hline 
$29 \, {\rm T}$ &  \, $\,16.6  \, {\rm K}\,$& \, $16.2 \,  {\rm K}  \,$  \\\hline 
\end{tabular} 
\caption[Tab. S]{Transition temperature $(T_{c})$  from PM to low temperature magnetic state at various applied fields. 
$T_{s}$  marks the onset temperature for the appearance of the local staggered field. 
This table contains $T_{s}$ values deduced from all of our NMR measurements, including 
our data presented here and that obtained previously in different magnetic fields from those  presented in \mbox{Figures. \ref{Fig1}, \ref{Fig2}, \& \ref{Fig3}}.
 }
\label{Table1}
\end{center}
\vspace*{-0.5cm}
\end{table}

$T_{c}$ from  PM to low temperature FM state was determined 
   by the crossing points of the plotted solid lines in  \mbox{Fig. \ref{Fig2}}.  Evidently, transition temperature increases with increasing applied field, as summarized  in \mbox{Tab. \ref{Table1}}. The increase of $T_{c}$ is significant and approximately  scales as the magnetic energy associated with the applied field,  confirming  the  magnetic nature of the transition. 
   Furthermore, we determine $T_{s}$  from   the onset temperature  which indicates the appearance of the local staggered field. As expected, the field dependence of $T_{s}$ closely follows that of $T_{c}$. This establishes that the 
rise of canted moments coincides with the development of LRO magnetism. 

 \begin{center}	
 %
\begin{figure}[h]
  \vspace*{-0.0cm}
\begin{minipage}{0.98\hsize}
 \centerline{\includegraphics[scale=0.95]{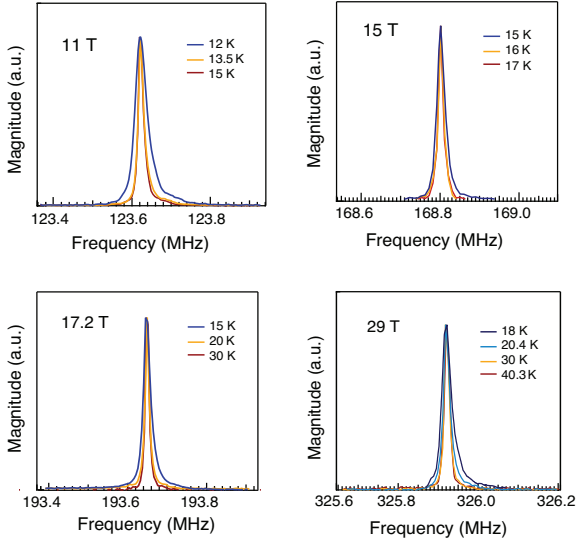}} 
\begin{minipage}{.98\hsize}
 \vspace*{-0.0cm}
\caption[]{\label{Fig3} \small 
Temperature dependence of the \Na NMR spectra measured in the PM phase at applied magnetic fields ranging from 11 T to 29 T, as denoted. This is a subset of data plotted in \mbox{Fig. \ref{Fig1}}, 
to illustrate the appearance of notable line broadening on decreasing temperature.  
  }
 \vspace*{-0.3cm}
\end{minipage}
\end{minipage}
\end{figure}
%
\end{center}
  \vspace*{-0.90cm}
%

\begin{center}	
 %
\begin{figure}[t]
  \vspace*{-0.0cm}
\begin{minipage}{0.98\hsize}
 \centerline{\includegraphics[scale=0.55]{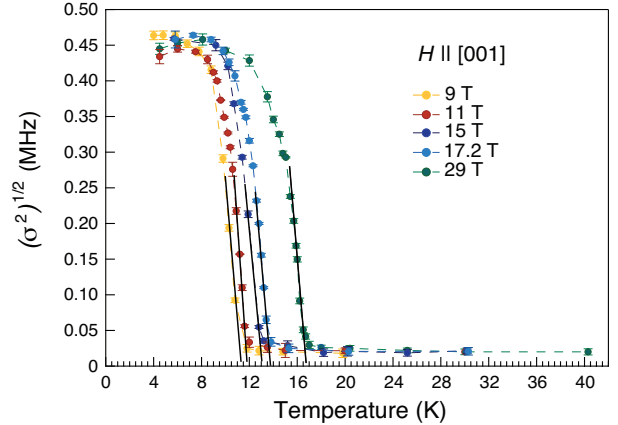}} 
\begin{minipage}{.98\hsize}
 \vspace*{-0.0cm}
\caption[]{\label{Fig4} \small 
Temperature dependence of the  second moment of the NMR spectral line, measuring the spectral width,  at applied magnetic fields ranging from 9 T to 29 T, as denoted. 
Solid lines are linear fits to the data  while dashed lines are guide to the eyes. 
  }
 \vspace*{-0.3cm}
\end{minipage}
\end{minipage}
\end{figure}
%
\end{center}
  \vspace*{-0.90cm}
%

\subsection{Intermediate  Temperature - BLPS Phase}

\begin{table}[h]
\begin{center}
\begin{tabular}{|c|c|c|}\hline $\quad H \, \|  [001] \quad$ & \, $T_{c}$  & \, $T^{*}$   \\\hline 
 $0 \, {\rm T}$ & \, $\,6.3 \,  {\rm K}\,$ \cite{{Erickson07}}& \,      
  \\\hline $7 \, {\rm T}$ & \, $\,9.9 \,  {\rm K}\,$& \, $11.5 \,  {\rm K}  \,$     \\\hline 
$9 \, {\rm T}$ &  \,  $\,10.8\, {\rm K}\,$& \, $12 \,  {\rm K}  \,$    \\\hline 
$11 \, {\rm T}$ &  \, $\,11.6  \, {\rm K}\,$& \, $13 \,  {\rm K}  \,$  \\\hline 
$15 \, {\rm T}$ &  \, $\,13   \, {\rm K}\,$& \, $15 \,  {\rm K}  \,$  \\\hline 
$17.2 \, {\rm T}$ &  \, $\,13.6  \, {\rm K}\,$& \, $16 \,  {\rm K}  \,$  \\\hline 
$29 \, {\rm T}$ &  \, $\,16.6  \, {\rm K}\,$& \, $20.4 \,  {\rm K}  \,$  \\\hline 
\end{tabular} 
\caption[Tab. S]{Transition temperature $(T_{c})$  from PM to low temperature magnetic state at various applied fields. $T^{*}$ denotes the onset temperature for    breaking of local cubic symmetry.  This table contains $T_{s}$ and $T^{*}$ values deduced from all of our NMR measurements, including 
our data presented here and that obtained previously in different magnetic fields from those  presented in \mbox{Figures. \ref{Fig1}, \ref{Fig2}, \& \ref{Fig3}}.}
\label{Table2}
\end{center}
\vspace*{-0.5cm}
\end{table}

     \begin{center}	
 %
\begin{figure}[t]
  \vspace*{-0.0cm}
\begin{minipage}{0.98\hsize}
 \centerline{\includegraphics[scale=0.55]{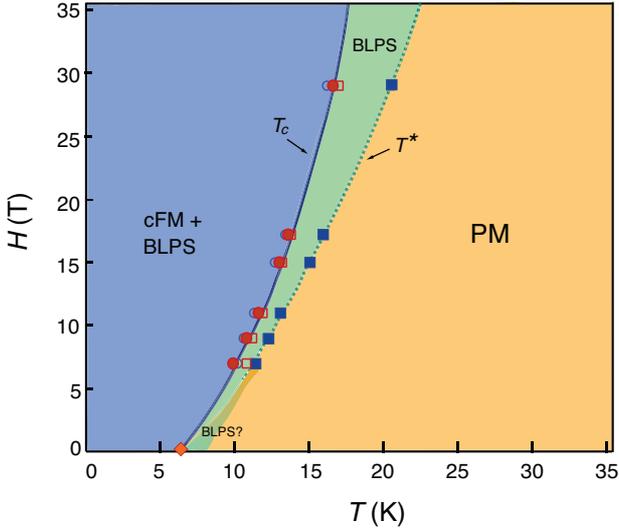}} 
\begin{minipage}{.98\hsize}
 \vspace*{-0.0cm}
\caption[]{\label{Fig5} \small 
Sketch of the phase diagram based on our NMR measurements, both presented here and in \mbox{Ref. \cite{Lu17}}.  Squares indicate onset  temperature   for the local cubic symmetry breaking,     determined from our NMR data as explained in the text, in the paramagnetic (PM)  phase. Solid circles denote $T_{\rm c}$, transition temperature into canted ferromagnetic (cFM) phase, as deduced from the NMR data. Diamond mark $T_{\rm c}$ as determined from thermodynamic measurements in \mbox{Ref. \cite{Erickson07}}. Open circles denote  onset temperature for the appearance of $H_{{stag}}$, while open squares mark  onset temperature for rapid increase of the NMR spectral line broadening (see  \mbox{Fig. \ref{Fig4}}).  
The solid line  indicates phase transition into cFM state and also 
possible tetragonal-to-orthorhombic  phase transition \cite{Lu17}. The dashed  line denotes cross over to the   BLPS phase, as illustrated in  \mbox{Fig. \ref{Fig3}}.}
 \vspace*{-0.3cm}
\end{minipage}
\end{minipage}
\end{figure}
%
\end{center}
  \vspace*{-0.90cm}
%
 %
%

Next, we proceed to identify the onset temperature for breaking of local cubic symmetry $(T^{*})$.
 Breaking of local cubic symmetry is accompanied by the appearance of finite EFG at the Na site. 
Non-zero EFG splits the single \Na line into three lines. However, for 
small finite values of the EFG, significant line broadening can only be detected \cite{Lu17}. 
Therefore, we identify $(T^{*})$ as  the temperature below which the width of the NMR spectral line 
  increases  notably   compared to that  in  high temperature PM phase. The procedure is illustrated in \mbox{Fig. \ref{Fig3}}. Deduced values of $T^{*}$ are listed in \mbox{Tab. \ref{Table2}}. 
  Breaking of local cubic symmetry indicates the appearance of   possible  orbital and/or quadrupolar ordering \cite{ChenBalents10, KrempaRev14}. 
  We point out that $T^{*}$ does not necessarily correspond to the true transition temperature from PM cubic to orbitally ordered  BLPS  state. This is because this transition could be   undetectable in our experiment due to the subtlety of the effect.

 The second moment of the NMR spectral line represents a quantitative  measure  of the  width of the NMR spectral line. 
In \mbox{Fig. \ref{Fig4}} we plot the temperature evolution of the second moment of the NMR spectra at various applied fields. On lowering $T$,  the second moment is dominated by the increase associated with  the appearance of both triplet satellite transitions  and magnetism. 
Therefore, the temperature corresponding to the onset of  rapid increase of the second moment of the NMR line,  determined  by the crossing points of the plotted solid lines and x-axis, does not correspond to $T^{*}$ but is dominated by magnetic broadening contribution.

 \section{Phase Diagram of \BaOs}
  A phase diagram   determined by our NMR measurements is plotted in  \mbox{Fig. \ref{Fig5}}. 
As magnetic field increases, BLPS phase extends over a larger temperature range.   

\section{Summary}
In conclusion, we   measured the \Na NMR  spectra  of   a single crystal sample of  \BaOsS  as a function of    temperature in different magnetic fields, applied parallel to [001] direction. 

Clear  NMR signatures of   phase transition to LRO magnetic phase is   identified. The  transition is  characterized by the sizable discontinuity of the local uniform field and abrupt appearance of  the local staggered field.  In addition, the  onset temperature for breaking of local cubic symmetry is determined. 
We find that BLPS phase occupies  a larger portion of the phase diagram in high fields.

   \section{Acknowledgement}
 The study was supported in part by the the National Science Foundation  DMR-1608760.       The study at the NHMFL was supported by the National Science Foundation under Cooperative Agreement no. DMR95-27035,  the State of Florida, and Brown University. 
   Work at Stanford University was  supported by the DOE, Office of Basic Energy Sciences, under Contract No.  DE-AC02-76SF00515. 




\end{document}